\documentclass[%
  prd,
  amsmath,amssymb,
  aps,
  reprint,
]{revtex4-2}

\usepackage{graphicx}
\usepackage{dcolumn}
\usepackage{bm}

\newcommand{\fdg}{\hbox{$.\!\!^\circ$}}

\begin{document}

\title{Do Primordial Black Hole Clusters Survive the Galaxy? Collisional Disruption and Microlensing Implications}

\author{M.V. Tkachev}\email{mtkachev@asc.rssi.ru}
\author{S.V. Pilipenko}%
 \email{spilipenko@asc.rssi.ru}
 \affiliation{%
Astro Space Center of P.N. Lebedev Physical Institute, Moscow, Russia}

\date{\today}

\begin{abstract}
We study the collisional disruption of primordial black hole (PBH) clusters in the
Milky Way halo.  Encounters between clusters strip PBHs into a diffuse component,
and the fraction of PBH mass in this smooth component along a sightline determines
how microlensing constraints divide between isolated compact objects and extended
cluster lenses.  We combine an analytic NFW-based collision-rate model, $72$ direct
N-body binary collision simulations that calibrate the escaped-mass fraction as a
universal function $\tilde f(\tilde v,\tilde b)$ of the relative velocity and
impact parameter in units of the cluster velocity scale and half-mass radius, and cosmological N-body simulations of a Milky Way-like halo
($M_{200}\simeq 8\times10^{11}\,M_\odot$, $c_{200}\simeq 11$) that record the
encounter history of every cluster from $z=9$ to $z=0$.  For $10^6$ and
$10^7\,M_\odot$ clusters --- bracketing the maximum mass in the Carr et al.\
formation scenario --- the local encounter rate at the Solar circle is
$2.9\times10^{-3}$ and $1.2\times10^{-2}\,\rm Myr^{-1}$, consistent with the
simulations to within ${\sim}40\%$.  Because the peak-disruption velocity of
Carr-radius clusters ($12$--$20\,\rm km\,s^{-1}$) lies far below typical halo
encounter velocities, disruption accumulates through many weak encounters, most
effectively during the early, cold phases of halo assembly: half of the total mass
loss is inflicted before $z\approx2$, a channel that $z=0$ analytic estimates miss
entirely.  The surviving mass fraction at the Solar circle is $S\simeq0.50$
($10^6\,M_\odot$) and $0.04$ ($10^7\,M_\odot$), and the DM-mass-weighted smooth
fraction toward the LMC and SMC is $0.49$ and $0.92$, respectively.
Cluster-cluster disruption is thus substantial over the Galaxy's lifetime, and
reanalyses of microlensing surveys must account for the radially varying smooth
fraction $f_{\rm sm}(r)$ derived here.
\end{abstract}

\maketitle

\section{Introduction}

The nature of dark matter (DM) remains one of the outstanding problems in modern
cosmology.  Among the candidates that have attracted renewed attention is the possibility
that a fraction of the DM consists of Primordial Black Holes (PBHs) formed in the
radiation-dominated era through the collapse of large overdensities
\citep[e.g.,][]{ZeldovichNovikov1967, Hawking1971, Carr1974, Chapline1975, Ivanov1994}.  Observational constraints on the compact-object fraction of DM now span multiple
independent probes: gravitational microlensing surveys by EROS, OGLE, and Subaru Hyper
Suprime-Cam (HSC) \citep{EROS2007, Mroz2024, Niikura2019}, disruption of wide stellar
binaries \citep{Shariat2025}, and gravitational-wave merger-rate analyses from
LIGO/Virgo/KAGRA \citep{Carcasona2026}.  Together they cover a broad mass range, with
the stellar-mass window ($\sim 1$--$100\,M_\odot$) currently among the most actively
discussed.

A further complication in interpreting these constraints arises if PBHs reside not as
isolated compact objects but in gravitationally bound clusters.  Several formation
scenarios predict that Poisson fluctuations in the PBH number density seed cluster-mass
halos that collapse before the standard $\Lambda$CDM halo population
\citep{Clesse2015, Clesse2017, Clesse2018, Inman2019, Carr2019, DeLuca2020, Carr2024, Siles2025, Hawkins2025}.  We take the existence of such clusters as
a working hypothesis and focus on a question that is independent of the formation
mechanism: can PBH clusters survive the history of gravitational encounters accumulated
over the life of the Galaxy?

The existence of such clusters might alter microlensing predictions.  On one hand,
within a cluster the effective lensing cross-section corresponds to the cluster's Einstein
radius rather than that of a single PBH, shifting event timescales toward longer values
and modifying the optical depth.  If clusters are compact enough to act as single lenses,
their total mass shifts the detection threshold relative to the isolated-PBH case
\citep{Clesse2018}; even in more diffuse configurations, a significant fraction of PBH
mass produces light-curve profiles inconsistent with the point-lens point-source template
used by survey pipelines, making them escape detection \citep{Gorton2022, Petac2022, Toshchenko2025}.
On the other hand, clusters
undergo gravitational encounters with each other as they orbit the Milky Way halo.
These encounters transfer kinetic energy to individual PBHs, which may then escape the
cluster potential, gradually building up a smooth, unclustered component.  The ratio
of clustered to smooth PBH mass along any given line of sight determines whether
microlensing constraints based on individual compact objects are appropriate or must be
corrected for the clustered geometry.

Previous analytic estimates of cluster disruption have relied on simplified encounter
models and order-of-magnitude cross-sections \citep[e.g.,][]{Jedamzik2020, Franciolini2022}.
In this work, we pursue a fully calibrated approach.  We first derive an analytic
collision-rate model for clusters in an NFW halo, incorporating gravitational focusing
and the local velocity distribution, then run a systematic suite of $72$ direct N-body
binary cluster collision simulations to measure the fraction of cluster mass lost per
encounter.  Exploiting the scale invariance of Newtonian gravity, we cast this
calibration as a universal dimensionless function $\tilde f(\tilde v,\tilde b)$ of the
relative velocity in units of the cluster internal velocity scale,
$\tilde v = v_{\rm rel}/v_c$, and the impact parameter in units of the half-mass
radius, $\tilde b = b/r_h$, applicable to clusters of any mass and size that share the
same density profile shape.  The encounter
history is accumulated by running cosmological N-body simulations of a Milky Way-like
halo --- selected from a parent cosmological box by mass, concentration, and
isolation, and traced from $z=9$ to $z=0$ --- in which each
simulation particle represents a single PBH cluster.  Resolving a galaxy-scale halo down
to $30\,M_\odot$ objects is computationally out of reach, so the binary collision grid
provides the sub-grid mass-loss calibration for each encounter recorded by the halo
simulation.  From these two simulation sets we derive the radial cluster survival profile
and its projection along sightlines toward the Magellanic Clouds.

The paper is structured as follows.  Section~\ref{sec:setup} describes the NFW halo
model and cluster properties.  Section~\ref{sec:analytic} presents the analytic
collision-rate model.  Section~\ref{sec:nbody} details the N-body simulation suite.
Section~\ref{sec:massloss} presents the results of the binary-collision mass-loss and the
fitting function.  Section~\ref{sec:survival} derives the cumulative survival fraction
in the halo.  Section~\ref{sec:los} presents the line-of-sight smooth fractions toward
the Magellanic Clouds (LMC and SMC).  Section~\ref{sec:discussion} discusses caveats and broader
implications, and Section~\ref{sec:conclusions} summarizes our conclusions.

\section{Halo and Cluster Model}
\label{sec:setup}

\subsection{NFW Dark Matter Halo}
\label{sec:nfw}

We model the Milky Way (MW) dark matter halo as a spherically symmetric
Navarro-Frenk-White (NFW) profile \citep{NFW1997},
\begin{equation}
  \rho_{\rm DM}(r) = \frac{\rho_s}{x(1+x)^2}, \qquad x \equiv \frac{r}{r_s},
  \label{eq:nfw}
\end{equation}
where $\rho_s$ is the characteristic density and $r_s$ the scale radius.  Given a halo
mass $M_{200}$ and concentration $c_{200}$, the NFW parameters follow from
\begin{equation}
\begin{aligned}
  R_{200} &= \left[\frac{3\,M_{200}}{4\pi\cdot 200\,\rho_{c,0}}\right]^{1/3},
  \qquad r_s = \frac{R_{200}}{c_{200}},\\
  \rho_s &= \frac{200\,\rho_{c,0}}{3}\,\frac{c_{200}^3}{f(c_{200})},
\end{aligned}
\end{equation}
where $f(c)\equiv\ln(1+c)-c/(1+c)$ and $\rho_{c,0}=3H_0^2/(8\pi G)$ is the critical
density evaluated with $H_0=67.77\,\rm km\,s^{-1}\,Mpc^{-1}$.  The NFW parameters are
determined from the $z=0$ snapshot of the Milky Way-like N-body halo simulation
(Section~\ref{sec:ic_halo}) by fitting the spherically averaged density profile
(shrinking-sphere center, log-log least squares over $r=2$--$200$ kpc); for the
$10^6\,M_\odot$ run we find
\begin{equation}
\begin{aligned}
  \rho_s &= 6.8\times10^{-3}\,M_\odot\,{\rm pc^{-3}}, &
  r_s    &= 18.1\,{\rm kpc},\\
  R_{200}&= 194\,{\rm kpc}, &
  c_{200}&= 10.8,
\end{aligned}
  \label{eq:nfw_params}
\end{equation}
corresponding to a halo mass $M_{200}\simeq 7.8\times10^{11}\,M_\odot$ (the
$10^7\,M_\odot$ run, evolved from the same initial patch at lower resolution, gives
$c_{200}=10.1$ and $M_{200}\simeq 7.7\times10^{11}\,M_\odot$; each run's own fit is
used in its analysis).  The concentration matches Milky Way estimates of
$c_{200}\approx 10$--$15$ from stellar kinematics and satellite dynamics, while the
mass lies at the lower end of the observationally allowed range
\citep{McMillan2017, BlandHawthorn2016}.  The DM-only circular speed at the Solar
circle is $V_c(8.2\,{\rm kpc})\simeq128\,\rm km\,s^{-1}$.

The enclosed mass and circular speed are
\begin{equation}
  M(<r) = 4\pi\rho_s r_s^3\,f(r/r_s),
  \qquad
  V_c(r) = \sqrt{\frac{G\,M(<r)}{r}}.
\end{equation}

\subsection{PBH Cluster Properties}
\label{sec:clusters}

We consider two cluster mass scales, $M_{\rm cl} = 10^6\,M_\odot$ and $M_{\rm cl} =
10^7\,M_\odot$.  These bracket the maximum cluster mass expected in the Carr et al.\
\citep{Carr2024} formation scenario: fluctuations below that scale collapse into
clusters, so most PBHs end up clustered, and the upper mass cutoff is
$\sim10^6$--$10^7\,M_\odot$ \citep[eq.~II.13]{Carr2024}.  We adopt individual PBH
masses $m_{\rm PBH} = 30\,M_\odot$, so that a cluster contains
$N_{\rm PBH} = M_{\rm cl}/m_{\rm PBH} \simeq 3.3\times10^4$ or
$3.3\times10^5$ black holes, respectively.

We characterize the internal structure of a cluster by its half-mass radius $r_h$ and
the associated internal velocity scale
\begin{equation}
  v_c \equiv \sqrt{\frac{G M_{\rm cl}}{r_h}},
  \label{eq:vc}
\end{equation}
which sets the magnitude of internal motions and, as shown in
Section~\ref{sec:massloss}, the relative velocity at which cluster-cluster collisions
are most destructive.  We set $r_h$ from the PBH cluster formation model of \citet{Carr2024}: Poisson
fluctuations in the PBH number density seed overdensities $\delta \propto (M_{\rm
cl}/m_{\rm PBH})^{-1/2}$, which collapse to virialized clusters at density $178\times$
the background at the collapse redshift.  Because larger clusters have smaller initial
overdensities, they collapse later at lower background density; substituting the
collapse redshift from \citeauthor{Carr2024} eq.\,(II.11) into the virial condition
gives a formation radius
\begin{equation}
  r_h \simeq 135\,{\rm pc}
    \left(\frac{m_{\rm PBH}\,f_{\rm PBH}}{M_\odot}\right)^{-1/2}
    \!\left(\frac{M_{\rm cl}}{10^6\,M_\odot}\right)^{5/6}.
  \label{eq:rh}
\end{equation}
The $M_{\rm cl}^{5/6}$ scaling is steeper than the $M_{\rm cl}^{1/3}$ of a
constant-density model precisely because the background density at collapse decreases
with cluster mass.  For $m_{\rm PBH}=30\,M_\odot$ and $f_{\rm PBH}=1$ this yields
\begin{equation*}
  r_h \simeq 25\,{\rm pc}\quad(M_{\rm cl}=10^6\,M_\odot),
\end{equation*}
\begin{equation*}
  r_h \simeq 168\,{\rm pc}\quad(M_{\rm cl}=10^7\,M_\odot).
\end{equation*}
The corresponding internal velocity scales (Eq.~\ref{eq:vc}) are
$v_c \simeq 13.1\,\rm km\,s^{-1}$ and $v_c \simeq 16.0\,\rm km\,s^{-1}$, respectively:
under the Carr scaling $r_h\propto M_{\rm cl}^{5/6}$, $v_c$ grows only as
$M_{\rm cl}^{1/12}$.  For the N-body collision calibration we adopt a Dehnen density
profile with inner slope $\gamma=3/2$ \citep{Dehnen1993} (Section~\ref{sec:ic_single}).

The number density of clusters in the halo is
\begin{equation}
  n_{\rm cl}(r) = f_{\rm cl}\,\frac{\rho_{\rm DM}(r)}{M_{\rm cl}},
  \label{eq:ncl}
\end{equation}
where $f_{\rm cl}\in[0,1]$ is the fraction of the total DM mass contained in clusters.
We present results for $f_{\rm cl}=1$ (all DM in clusters) and note that rates scale
as $\Gamma\propto f_{\rm cl}^2/f_{\rm cl} = f_{\rm cl}$ when both the target and
projectile densities are proportional to $f_{\rm cl}$.

\section{Analytic Collision Rate Model}
\label{sec:analytic}

\subsection{Gravitational Focusing Cross-Section}

The effective cross-section for a collision between two clusters of equal mass $M_{\rm cl}$
with relative velocity $v_{\rm rel}$ and maximum impact parameter $b_{\rm cut}$ is
enhanced by gravitational focusing \citep{Spitzer1987},
\begin{equation}
  \sigma(b_{\rm cut},\,v_{\rm rel}) = \pi b_{\rm cut}^2\left[1 + \frac{v_{\rm esc}^2(b_{\rm cut})}{v_{\rm rel}^2}\right],
  \label{eq:sigma}
\end{equation}
where the pericentric escape speed for two equal-mass clusters is
$v_{\rm esc}^2(b_{\rm cut}) = 4GM_{\rm cl}/b_{\rm cut}$.  We identify $b_{\rm cut}$ with
the cutoff impact parameter beyond which a collision causes negligible internal
disruption, setting
\begin{equation}
  b_{\rm cut} = k\,r_h,
  \label{eq:bcut}
\end{equation}
with $k\sim 1$--$2$.  The binary N-body calibration described in
Section~\ref{sec:massloss} shows that the mass loss per encounter stays within
$\sim5\%$ of its head-on value out to $b=2r_h$ and declines on a characteristic
scale $b\simeq 7\,r_h$; the range $k=1$--$2$ therefore counts the encounters that
individually remove the most mass, and we treat the factor-of-two spread in $k$ as
part of the model uncertainty band.

We assume the cluster population is in approximate dynamical equilibrium within the NFW
potential, so that the one-dimensional velocity dispersion at radius $r$ obeys the
Jeans closure
\begin{equation}
  \sigma_{1D}(r) \simeq \frac{V_c(r)}{\sqrt{2}}.
  \label{eq:sigma1d}
\end{equation}
For a Maxwellian distribution of three-dimensional velocities with this dispersion, the
mean relative speed between two clusters is
\begin{equation}
  \langle v_{\rm rel}\rangle(r) = \frac{4}{\sqrt{\pi}}\,\sigma_{1D}(r),
  \label{eq:vrel}
\end{equation}
and the mean squared relative speed entering the cross-section denominator is
\begin{equation}
  v_{\rm eff}^2(r) \equiv \langle v_{\rm rel}^2\rangle(r) = 6\,\sigma_{1D}^2(r).
  \label{eq:veff}
\end{equation}
We use $v_{\rm eff}$ as the representative velocity in Equation~(\ref{eq:sigma}).

\subsection{Local Collision Rate}

The mean-free-path rate for a single cluster at radius $r$ to undergo a collision is
\begin{equation}
  \Gamma(r) = n_{\rm cl}(r)\,\sigma\!\left(b_{\rm cut},\,v_{\rm eff}(r)\right)\,\langle v_{\rm rel}\rangle(r),
  \label{eq:gammalocal}
\end{equation}
with $n_{\rm cl}(r)$, $\sigma$, and $\langle v_{\rm rel}\rangle$ given by
Equations~(\ref{eq:ncl})--(\ref{eq:veff}).  For the MW Solar neighbourhood
($r_\odot=8.2\,\rm kpc$), $k=1$, $f_{\rm cl}=1$, and the NFW parameters of
Equation~(\ref{eq:nfw_params}), we find
\begin{equation}
  \Gamma(r_\odot) \simeq
  \begin{cases}
    2.9\times10^{-3}\,\rm Myr^{-1} & (10^6\,M_\odot,\ \tau_{\rm coll}\simeq 350\,\rm Myr), \\
    1.2\times10^{-2}\,\rm Myr^{-1} & (10^7\,M_\odot,\ \tau_{\rm coll}\simeq 82\,\rm Myr).
  \end{cases}
  \label{eq:gammanumerical}
\end{equation}
Both collision times are far shorter than the Hubble time.  The rate rises steeply
toward the Galactic center because of the increasing NFW density,
and falls off at $r\gtrsim 50\,\rm kpc$.

\subsection{Halo-Averaged Rate}

To characterize the global collision activity, we compute the halo-averaged per-cluster
rate,
\begin{equation}
  \bar\Gamma = \frac{f_{\rm cl}}{M_{\rm cl}}\,\frac{\displaystyle\int_0^{R_{200}}
  \rho_{\rm DM}^2\,\sigma\!\left(v_{\rm eff}\right)\langle v_{\rm rel}\rangle\, r^2\,dr}
  {\displaystyle\int_0^{R_{200}}\rho_{\rm DM}\, r^2\,dr},
  \label{eq:gammabar}
\end{equation}
with all radial profiles evaluated at radius $r$ (the $4\pi$ factors cancel between
numerator and denominator).
Numerically, Equation~(\ref{eq:gammabar}) is evaluated on a logarithmically spaced radial
grid of $2048$ points from $10^{-2}r_s$ to $R_{200}$.  For the parameters above we
obtain $\bar\Gamma \simeq 4.8\times10^{-4}\,\rm Myr^{-1}$ ($10^6\,M_\odot$ case;
$1.9\times10^{-3}\,\rm Myr^{-1}$ for $10^7\,M_\odot$), reflecting the mass-weighting
that dilutes the contribution of the central, high-rate region.

\begin{figure*}[htbp]
  \centering
  \includegraphics[width=\linewidth]{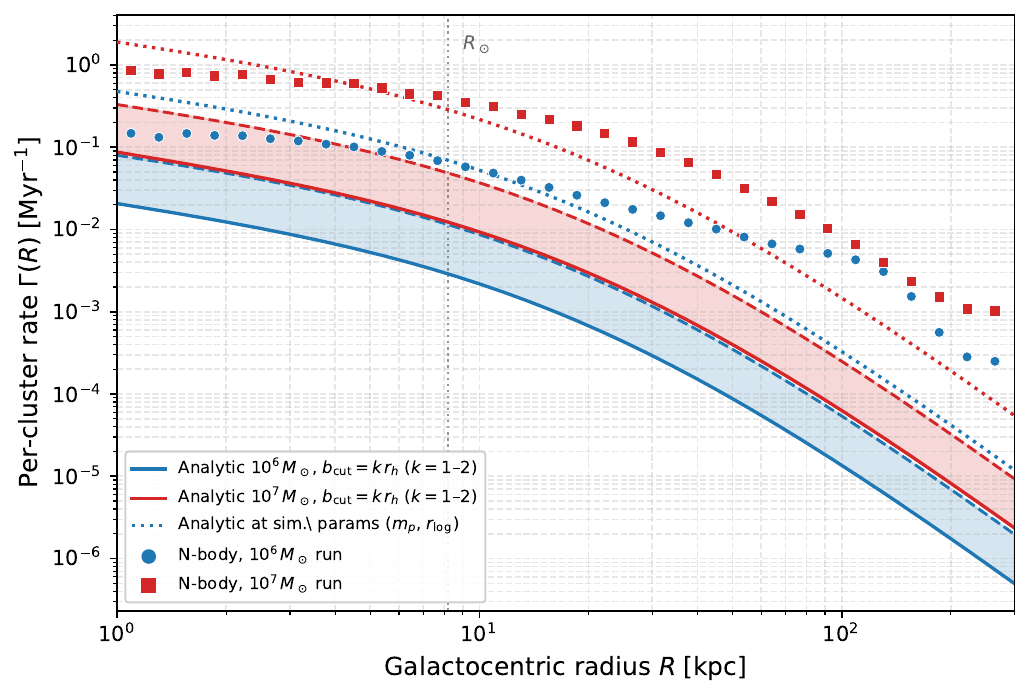}
  \caption{Radial profile of the local per-cluster encounter rate $\Gamma(r)$ for
    $M_{\rm cl}=10^6\,M_\odot$ (blue) and $10^7\,M_\odot$ (red) clusters
    ($f_{\rm cl}=1$; $r_h=25\,\rm pc$ and $168\,\rm pc$ from Eq.~\ref{eq:rh}),
    using the NFW parameters measured from each halo simulation.
    Shaded bands: analytic model for the Carr clusters with $b_{\rm cut}=k\,r_h$, spanning
    $k=1$ (solid) to $k=2$ (dashed).  Dotted lines: the same analytic model
    evaluated at each simulation's operational parameters (particle mass $m_p$ and
    encounter-logging radius $r_{\log}$; Section~\ref{sec:halo_sims}), which is the
    prediction the N-body points should be compared against.  Points: encounter rate per
    cluster measured from the halo simulations in radial bins over the late-time
    window $z\lesssim0.2$.  At the Solar circle (vertical dotted line,
    $r_\odot=8.2\,\rm kpc$) the simulation agrees with the analytic prediction to
    within $\sim40\%$; the deficit in the innermost few kpc and the excess in the
    outskirts are discussed in Section~\ref{sec:discussion}.}
  \label{fig:gamma_r}
\end{figure*}

We show in Figure~\ref{fig:gamma_r} the radial profile $\Gamma(r)$ for both cluster
mass scales, $M_{\rm cl}\in\{10^6,\,10^7\}\,M_\odot$.  At fixed $f_{\rm cl}$, heavier
clusters have lower number density but larger cross-sections; the net dependence is
$\Gamma\propto M_{\rm cl}^{2/3}$ in the geometric limit (Carr scaling; see Section~\ref{sec:discussion}), steepened further by gravitational
focusing in the inner halo.

\section{N-body Simulations}
\label{sec:nbody}

\subsection{Single-Cluster Initial Conditions}
\label{sec:ic_single}

Equilibrium initial conditions for the reference cluster used in the binary-collision
suite are generated with Agama
\citep{Vasiliev2019}\footnote{\texttt{https://github.com/GalacticDynamics-Oxford/AGAMA}}
as an isotropic Dehnen sphere with inner slope $\gamma=3/2$ \citep{Dehnen1993},
sampled with $N_{\rm PBH}=33{,}333$ particles of mass $m_{\rm PBH}=30\,M_\odot$
($M_{\rm cl}=10^6\,M_\odot$).  The realized half-mass radius, measured directly from
the particle sample, is $r_{50}=1.83\,\rm pc$, corresponding to an internal velocity
scale $v_c=48.5\,\rm km\,s^{-1}$ (Eq.~\ref{eq:vc}).  The reference cluster is thus
substantially more compact than the formation-model radius of
Equation~(\ref{eq:rh}); because Newtonian gravity is scale-free, the binary-collision
results are cast in dimensionless form and rescaled exactly to the Carr cluster
parameters (Section~\ref{sec:massloss}).  Neither cluster mass scale requires a
separate suite: the same universal calibration applies to the $10^6$ and
$10^7\,M_\odot$ clusters through their respective $v_c$ and $r_h$.

\subsection{Full-Halo Initial Conditions}
\label{sec:ic_halo}

The cluster population is embedded in a Milky Way-like halo drawn from a
cosmological $\Lambda$CDM parent simulation: a $64\,h^{-1}$ Mpc periodic box with
$512^3$ particles ($m_p=1.66\times10^{8}\,h^{-1} M_\odot$; $\Omega_m=0.307$,
$\Omega_\Lambda=0.693$, $h=0.6777$), evolved from $z=9$ to $z=0$.  MW-mass halo
candidates were shortlisted from the $z=0$ Rockstar \citep{Behroozi2013} catalog by
mass, concentration, and environment; we select a halo with catalog mass
$M_{200c}=8.7\times10^{11}\,h^{-1} M_\odot\approx1.3\times10^{12}\,M_\odot$ and
concentration $c_{200}\approx11.9$, matching the canonical MW values of
\citet{McMillan2017}.  Its nearest more massive neighbour lies $2.4\,h^{-1}$ Mpc
away and the nearest cluster-mass halo ($1.7\times10^{13}\,h^{-1} M_\odot$) at
$2.7\,h^{-1}$ Mpc, so the halo evolves in effective isolation.

The halo is re-simulated as an isolated Lagrangian patch: all particles within a
sphere around the halo at $z=0$ are identified, and their IDs are traced back to the
parent box's $z=9$ snapshot to obtain the patch's initial phase-space coordinates
($N_{\rm orig}=4407$ particles, total mass $7.3\times10^{11}\,h^{-1} M_\odot$).
Because the extraction sphere is finite, the re-simulated halo converges to a
slightly lower final mass than its parent-box counterpart
($M_{200}\simeq7.8\times10^{11}\,M_\odot$, Equation~\ref{eq:nfw_params}), while
retaining a MW-like concentration.

To reach particle masses equal to the cluster mass scales of interest, we upsample
the patch with a kernel density estimation (KDE) perturbation scheme: each new
particle selects a random parent from the original sample and inherits its position
and velocity, both perturbed by isotropic offsets with exponentially distributed
amplitudes scaled by the parent's $k=8$ nearest-neighbor distance in position and
velocity space, respectively (a single shared parent per new particle preserves the
local position--velocity correlation).  The patch is recentered on its center of
mass, the bulk velocity is subtracted, and the rare perturbation-tail outliers
($<10^{-4}$ of the mass) are dropped.  This yields
$N=733{,}683$ particles of $m_p=10^6\,h^{-1} M_\odot=1.48\times10^{6}\,M_\odot$ for
the $10^6\,M_\odot$ run (upsampling factor $246$) and $N=73{,}368$ particles of
$m_p=1.48\times10^{7}\,M_\odot$ for the $10^7\,M_\odot$ run (factor $25$).  The
resulting phase-space distributions are written as GADGET-2 IC files at $z=9$.  We
note that the $246\times$ upsampling substantially exceeds the factor of ${\sim}10$
at which the KDE scheme was originally validated; the density and velocity fields of
the upsampled patch were verified against the parent sample, but small-scale
clustering below the parent resolution is necessarily smooth.

\subsection{Binary Collision Grid}
\label{sec:ic_binary}

Binary collision ICs are constructed by placing two copies of the reference cluster
sample at a separation of $r_0 = 100\,\rm pc$ along the $x$-axis,
with the second cluster's center-of-mass boosted by a relative velocity
$\mathbf{v}_{\rm rel} = v_{\rm rel}(\cos\theta,\,\sin\theta,\,0)$ in the orbital plane,
where $\theta$ is the angle between the relative velocity vector and the
inter-cluster axis.  Because the clusters deflect little during an encounter
(Section~\ref{sec:massloss}), the relative trajectory is nearly rectilinear and the
encounter geometry is fully specified by the impact parameter
\begin{equation}
  b = r_0\,|\sin\theta|.
  \label{eq:impact}
\end{equation}

The suite comprises $72$ completed simulations spanning $17$ relative velocities from
$1$ to $908\,\rm km\,s^{-1}$ ($\tilde v = v_{\rm rel}/v_c \simeq 0.02$--$19$) and
impact parameters from $0$ to $100\,\rm pc$ ($\tilde b = b/r_{50}$ from $0$ to $55$):
a head-on ($b=0$) velocity scan, a dense impact-parameter scan at the peak-disruption
velocity $v_{\rm rel}\simeq74\,\rm km\,s^{-1}$ ($\tilde v\simeq1.5$), and coarser
angular grids at seven additional velocities.  For each run, the simulation
duration is determined by integrating the two-body equation of motion numerically to find
the time at which the cluster separation first exceeds $r_{\rm esc}=200\,\rm pc$ after
closest approach (or $8\,\rm Myr$ for bound pairs).  Individual GADGET-2 parameter
files are auto-generated with the appropriate \texttt{TimeMax} value.  All runs use a gravitational softening of
$\epsilon = 0.1\,\rm pc$.

\subsection{Halo Encounter Simulations}
\label{sec:halo_sims}

The full-halo simulations use \texttt{g2\_enc}%
\footnote{\texttt{https://github.com/tkachev-asc/g2\_enc}},
a modified version of GADGET-2 \citep{Springel2005} in which a compile-time
\texttt{ENCOUNTER\_LOGGER} module is added to the standard tree-gravity solver.
The simulations are dark-matter only; no baryonic (stellar or gas) component is
included.  Each simulation particle represents one PBH cluster of mass $m_p$; the
corresponding Carr formation radii and internal velocity scales
(Equations~\ref{eq:rh} and~\ref{eq:vc}) are $r_h=34\,\rm pc$,
$v_c=13.7\,\rm km\,s^{-1}$ for the $10^6\,M_\odot$ run and $r_h=232\,\rm pc$,
$v_c=16.5\,\rm km\,s^{-1}$ for the $10^7\,M_\odot$ run.  Because these radii are
much smaller than the inter-cluster mean spacing ($\sim$ kpc), clusters are treated
as point masses for orbital dynamics; their internal structure enters only via the
binary-collision calibration.

The encounter logger maintains a per-timestep open-addressing hash table that
deduplicates particle pairs.  Every pair whose physical separation falls below a
threshold $r_{\log}$ is flagged; each particle accumulates the number of unique
partners it encountered during the inter-snapshot interval, together with the
midpoint position $(\bar{x},\bar{y},\bar{z})$, the scalar relative speed
$\Delta v=|\mathbf{v}_1-\mathbf{v}_2|$, and the approach cosine
$\cos\theta=\hat{\mathbf{r}}\cdot\hat{\mathbf{v}}_{\rm rel}$ of its most recent
encounter.  Records are gathered
across MPI (Message Passing Interface) ranks and dumped to a per-snapshot binary
file at each output time.

We run two halo simulations from the same initial patch, both accumulating $51$
snapshots uniformly spaced in cosmic time ($\Delta t\simeq265\,\rm Myr$) from $z=9$
to $z=0$.  The $10^6\,M_\odot$ run uses a physical (proper) gravitational softening
$\epsilon=25\,h^{-1}\,{\rm pc}\approx37\,\rm pc$ and encounter-logging radius
$r_{\log}=100\,h^{-1}\,{\rm pc}\approx148\,\rm pc$; the
$10^7\,M_\odot$ run uses $\epsilon=168\,h^{-1}\,{\rm pc}\approx248\,\rm pc$ and
$r_{\log}=672\,h^{-1}\,{\rm pc}\approx992\,\rm pc$.  In both
cases $r_{\log}\simeq4.3\,r_h$, which captures essentially all disruptive
encounters: the impact-parameter suppression factor calibrated in
Section~\ref{sec:massloss} stays above $g\simeq0.77$ throughout the logged range.
More distant encounters are not recorded and therefore do not enter the mass-loss
accounting of Section~\ref{sec:survival}; the resulting bias is conservative and
is discussed in Section~\ref{sec:discussion}.

\section{Mass Loss from Binary Cluster Encounters}
\label{sec:massloss}

\subsection{Energy-Based Escape Criterion}

At the end of each binary collision simulation we identify which particles have escaped
by computing the total mechanical energy of each particle in the frame of the combined
system.  The cluster potential $\Phi(r)$ is calculated from the initial single-cluster
N-body snapshot and interpolated radially, and we define
\begin{equation}
  \Phi_{\rm tot}(\mathbf{r}) = \Phi_A(\mathbf{r}) + \Phi_B(\mathbf{r}),
  \label{eq:phi_tot}
\end{equation}
where $\Phi_A$ and $\Phi_B$ are the potentials of clusters A and B evaluated at the
particle position.  A particle is classified as unbound (escaped) if its energy in both
individual cluster frames and in the combined center-of-mass frame is positive,
\begin{equation}
  E_{A,i} > 0, \quad E_{B,i} > 0, \quad E_{{\rm cm},i} > 0,
  \label{eq:escape}
\end{equation}
where $E_{A,i} = \tfrac{1}{2}|\mathbf{v}_i-\mathbf{V}_A|^2 + \Phi_A(\mathbf{r}_i)$ and
similarly for $B$.  The escape fraction is then
\begin{equation}
  f_{\rm esc}(v_{\rm rel},\,b) = \frac{N_{\rm esc}}{N_{\rm tot}},
  \label{eq:fesc}
\end{equation}
where $N_{\rm tot}=2N_{\rm PBH}$ is the total initial number of particles.  The cluster
centers $\mathbf{R}_{A,B}$ are determined at each snapshot using an iterative shrinking-sphere
algorithm \citep{Power2003}.

\subsection{Universal Velocity Dependence}

Newtonian gravity contains no intrinsic scale: rescaling all lengths by a factor
$\lambda$ and all velocities by $\lambda^{-1/2}$ maps an equilibrium N-body system
onto another equilibrium system with the same mass and profile shape.  The escape
fraction produced by an encounter can therefore depend only on the dimensionless
combinations
\begin{equation}
  \tilde v \equiv \frac{v_{\rm rel}}{v_c}, \qquad
  \tilde b \equiv \frac{b}{r_{50}},
  \label{eq:dimensionless}
\end{equation}
with $v_c$ given by Equation~(\ref{eq:vc}); here $r_{50}$ denotes the measured
half-mass radius, identical to $r_h$ for the model clusters of
Section~\ref{sec:clusters}.  A single simulation suite of the
reference cluster thus calibrates a universal function $\tilde f(\tilde v,\tilde b)$
valid for all clusters sharing the same (Dehnen) profile shape, including both Carr
mass scales.

\begin{figure*}[htbp]
  \centering
  \includegraphics[width=\linewidth]{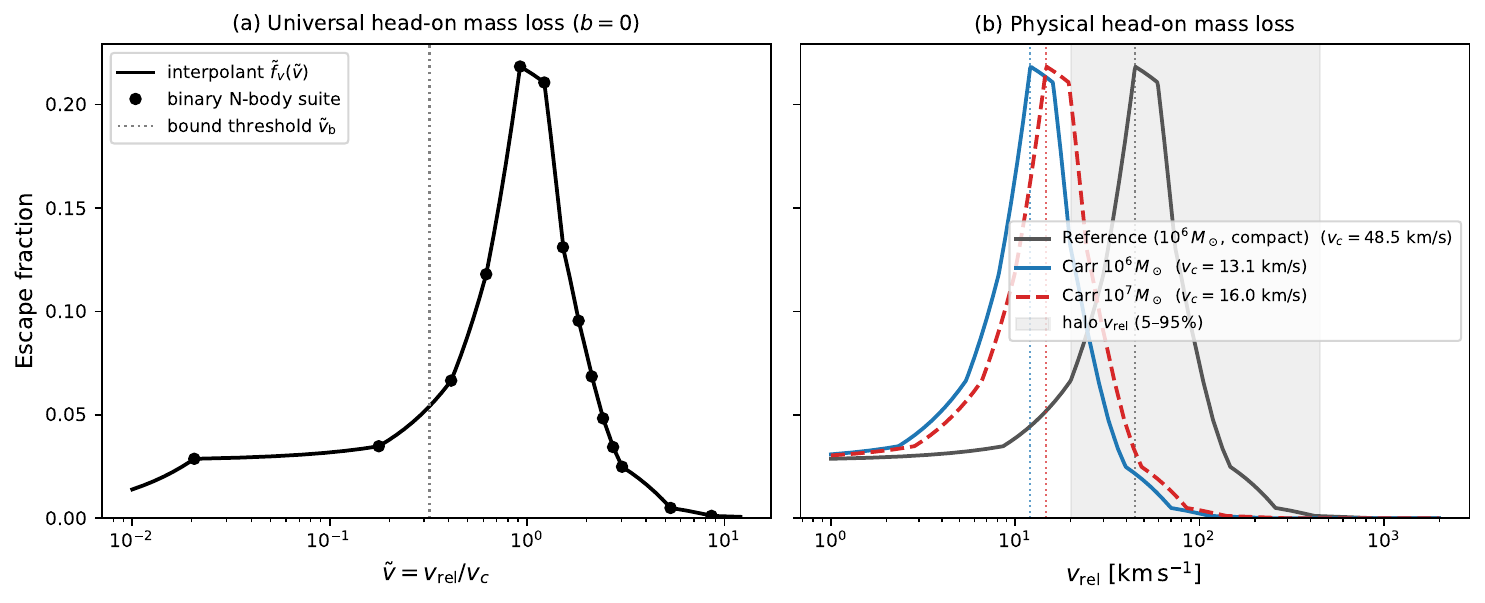}
  \caption{\textit{Left panel (a):} Universal head-on ($b=0$) escape fraction
    $\tilde f_v$ as a function of $\tilde v = v_{\rm rel}/v_c$.  Points are direct
    N-body measurements from the binary-collision suite (reference cluster,
    $r_{50}=1.83\,\rm pc$, $v_c=48.5\,\rm km\,s^{-1}$); the solid curve is the
    piecewise-linear interpolant $\tilde f_v(\tilde v)$.  The vertical dotted line
    marks the bound threshold $\tilde v_{\rm b}\approx 0.32$.
    \textit{Right panel (b):} the same calibration in physical units for the compact
    reference cluster (gray), Carr $10^6\,M_\odot$ clusters (blue solid), and Carr
    $10^7\,M_\odot$ clusters (red dashed).  Vertical dotted lines mark each cluster's
    peak-disruption velocity; the shaded band spans the 5th--95th percentile of the
    count-weighted encounter velocities measured in the halo simulations
    ($20$--$450\,\rm km\,s^{-1}$).}
  \label{fig:fv}
\end{figure*}

Figure~\ref{fig:fv}(a) shows the head-on escape fraction
$\tilde f_v(\tilde v)\equiv\tilde f(\tilde v,\,\tilde b=0)$.  The function exhibits a
clear peak, $\tilde f_v\approx 0.22$ at $\tilde v\simeq 0.9$--$1.2$: encounters are
most destructive when the relative velocity is comparable to the internal velocity
scale of the clusters.  The escape fraction declines toward higher velocities (fast
fly-bys transfer less energy in the impulsive approximation, $\Delta E \propto
v^{-2}$) and toward lower velocities, where the pair becomes gravitationally bound
($\tilde v < \tilde v_{\rm b} \approx 0.32$, comparable to the two-cluster escape
speed at the initial separation) and may merge rather than disrupt.  We construct a
piecewise-linear interpolant $\tilde f_v(\tilde v)$ directly from the head-on data,
anchored to zero at $\tilde v=0$ and beyond the last grid point.

Figure~\ref{fig:fv}(b) translates the universal curve into physical units.  For the
compact reference cluster the peak sits at $v_{\rm rel}\simeq 45$--$59\,\rm
km\,s^{-1}$, but for the Carr formation radii the peak disruption velocity is much
lower: $v_{\rm rel}\simeq 12$--$16\,\rm km\,s^{-1}$ for $10^6\,M_\odot$ clusters and
$\simeq 15$--$20\,\rm km\,s^{-1}$ for $10^7\,M_\odot$ clusters.  The encounter
velocities measured in the halo simulations (Section~\ref{sec:survival}) span
$v_{\rm rel}\simeq 20$--$450\,\rm km\,s^{-1}$ (count-weighted 5th--95th percentile,
median ${\simeq}230\,\rm km\,s^{-1}$), i.e.\ $\tilde v\simeq 1.5$--$32$ for the Carr
clusters.  The bulk of the encounters therefore falls far on the fast-fly-by side of
the peak, where the escape fraction per encounter is below $10^{-2}$; only the
low-velocity tail (${\sim}8\%$ of encounters with $\tilde v\lesssim3$ in the
$10^6\,M_\odot$ run, concentrated at early epochs) reaches into the
peak-disruption regime.

\subsection{Impact-Parameter Dependence}

\begin{figure}[htbp]
  \centering
  \includegraphics[width=\linewidth]{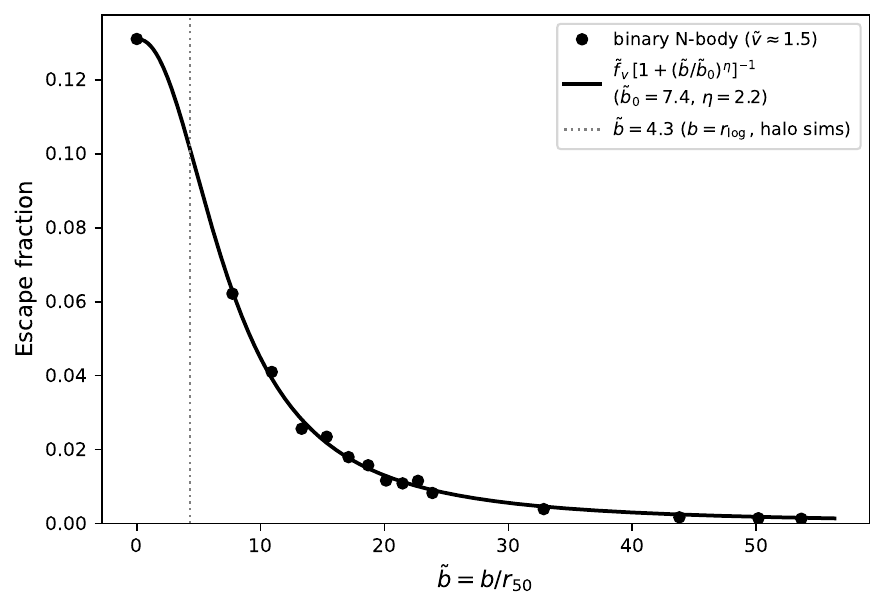}
  \caption{Universal impact-parameter dependence of the escape fraction at
    $\tilde v\approx 1.5$.  Points are binary N-body measurements with
    $b=r_0|\sin\theta|$ (Equation~\ref{eq:impact}); the curve is
    $\tilde f_v\,g(\tilde b)$ with the fitted suppression factor of
    Equation~(\ref{eq:gb}) ($\tilde b_0=7.4$, $\eta=2.2$).  The vertical dotted line
    marks $\tilde b=4.3$, corresponding to the encounter-logging radius
    $r_{\log}\simeq4.3\,r_h$ of the halo simulations.}
  \label{fig:fb}
\end{figure}

For fixed $\tilde v$, the escape fraction declines with increasing impact parameter:
head-on collisions ($b=0$) maximize the tidal impulse, while distant fly-bys perturb
only the cluster outskirts.  Figure~\ref{fig:fb} shows the impact-parameter scan at
the peak-region velocity $\tilde v\simeq 1.5$, with $b$ obtained from the run geometry
via Equation~(\ref{eq:impact}).  Note that approaching ($\cos\theta<0$) and separating
($\cos\theta>0$) initial configurations correspond to the same impact parameter
$b=r_0|\sin\theta|$, and the measured escape fractions are indeed symmetric between
the two --- consistent with the nearly rectilinear relative trajectories of these
extended, slowly deflecting systems.

We parametrize the suppression with
\begin{equation}
  g(\tilde b) = \left[1 + \left(\tilde b/\tilde b_0\right)^{\eta}\right]^{-1},
  \label{eq:gb}
\end{equation}
normalized to $g(0)=1$; a non-linear least-squares fit to the $\tilde v\simeq1.5$ data
gives $\tilde b_0=7.4$ and $\eta=2.2$.  Two features of this fit are worth noting.
First, $g$ declines slowly within the logging radius of the halo simulations
($b\le r_{\log}\simeq4.3\,r_h$, Section~\ref{sec:halo_sims}): $g(4.3)\approx0.77$,
and averaged over impact parameters (uniform in area) $\langle g\rangle\approx0.88$,
so logged encounters are on average nearly as damaging as head-on collisions.
Second, the mass loss extends well beyond the cluster
itself --- $g$ falls to one half only at $\tilde b\simeq 7$ --- reflecting the
effectiveness of tidal perturbations on the loosely bound cluster outskirts.

\subsection{Universal Model and Rescaling}

The combined mass-loss model is
\begin{equation}
  f(v_{\rm rel},\,b) = \tilde f_v\!\left(\frac{v_{\rm rel}}{v_c}\right)\,
  g\!\left(\frac{b}{r_h}\right),
  \label{eq:fcombined}
\end{equation}
exact for clusters of any mass and half-mass radius at fixed profile shape.  Below the
bound threshold ($\tilde v<\tilde v_{\rm b}$) the impact-parameter suppression is not
applied (factor set to unity) because bound interactions do not exhibit the impulsive
scaling.

Applied to the Carr formation radii, the relevant velocity scale follows from
Equations~(\ref{eq:vc}) and~(\ref{eq:rh}): with $r_h\propto M_{\rm cl}^{5/6}$,
\begin{equation}
  v_c \simeq 13.1\,{\rm km\,s^{-1}}
  \left(\frac{M_{\rm cl}}{10^6\,M_\odot}\right)^{1/12},
  \label{eq:vscale}
\end{equation}
so both Carr mass scales share nearly the same $v_c$
($13.1$ and $16.0\,\rm km\,s^{-1}$), an order of magnitude below the reference
cluster's $48.5\,\rm km\,s^{-1}$.  We verified the dimensionless rescaling directly
with a set of additional head-on collision simulations in which the reference sample
was rescaled to $r_h=25\,\rm pc$ (positions scaled by $\lambda\approx13.7$, velocities
by $\lambda^{-1/2}$; $v_{\rm rel}=1$--$500\,\rm km\,s^{-1}$): the measured escape
fractions follow the universal curve of Figure~\ref{fig:fv}(a), confirming in
particular the shift of the peak-disruption velocity to
$v_{\rm rel}\approx 12$--$16\,\rm km\,s^{-1}$.

\begin{figure*}[htbp]
  \centering
  \includegraphics[width=\linewidth]{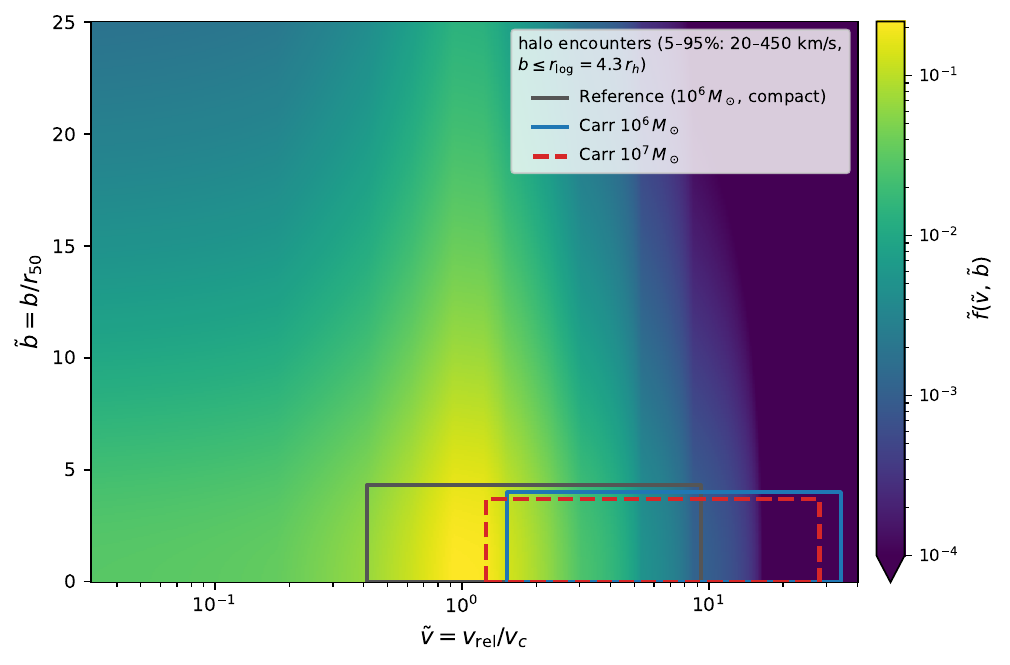}
  \caption{Universal escape fraction
    $\tilde f(\tilde v,\tilde b)=\tilde f_v(\tilde v)\,g(\tilde b)$
    (logarithmic color scale; values below $10^{-4}$ are shown at the color floor).
    Rectangles mark the region occupied by halo encounters
    ($v_{\rm rel}=20$--$450\,\rm km\,s^{-1}$, the 5th--95th percentile measured in
    the halo simulations; $b\le r_{\log}\simeq4.3\,r_h$; box heights staggered for
    visibility) for the compact reference cluster (gray), Carr $10^6\,M_\odot$
    clusters (blue), and Carr $10^7\,M_\odot$ clusters (red dashed).  The reference
    cluster's region covers the disruption peak at $\tilde v\approx 1$, whereas
    the bulk of the Carr clusters' halo encounters falls in the weak fast-fly-by
    regime.  The vertical dependence is weak within the encounter boxes by
    construction: the suppression scale $\tilde b_0=7.4$ lies above $r_{\log}$, so
    $g\ge0.77$ throughout the logged range and the mass loss there is controlled
    almost entirely by $\tilde v$.}
  \label{fig:fheatmap}
\end{figure*}

The full model $\tilde f(\tilde v,\tilde b)$ is shown in Figure~\ref{fig:fheatmap},
together with the region of parameter space occupied by halo encounters
($v_{\rm rel}=20$--$450\,\rm km\,s^{-1}$, $b\le r_{\log}\simeq4.3\,r_h$) for each
cluster type.  For the
compact reference cluster this region covers the disruption peak; for the Carr
clusters the bulk lies far to the fast side, where $\tilde f\lesssim 10^{-2}$,
with only the low-velocity edge touching the peak.  Averaged
over impact parameters $b\le r_{\log}$ (uniform in area), the mass loss per encounter
for Carr $10^6\,M_\odot$ clusters is $\langle f\rangle\simeq 1.6\times10^{-2}$ at
$v_{\rm rel}=50\,\rm km\,s^{-1}$, $2.0\times10^{-3}$ at $100\,\rm km\,s^{-1}$, and
below $10^{-3}$ for $v_{\rm rel}\gtrsim 150\,\rm km\,s^{-1}$; the $10^7\,M_\odot$
values are comparable ($2.1\times10^{-2}$, $3.5\times10^{-3}$, $\sim10^{-3}$).
Disruption of Carr clusters in the halo therefore proceeds not through rare
catastrophic collisions but through the cumulative effect of many weak encounters.

\section{Cumulative Survival Fraction in the Halo}
\label{sec:survival}

\subsection{Encounter Integration Method}

The halo simulations record, for each snapshot interval $i$ and each particle with at
least one encounter, the number of unique partners $c_i$ together with the midpoint
position, scalar relative speed $\Delta v_i$, and approach cosine $\cos\theta_i$ of
the particle's most recent encounter (Section~\ref{sec:halo_sims}).  Relative speeds
are converted from GADGET's internal velocity variable to physical peculiar
velocities, $\Delta v_{\rm phys}=\Delta v/a$, using the geometric-mean scale factor
of the interval.  Because a record is written when the pair separation first falls
below the logging radius, the (nearly rectilinear) encounter geometry gives the
impact parameter as in Equation~(\ref{eq:impact}) with $r_0\to r_{\log}$:
\begin{equation}
  b_i = r_{\log}\,|\sin\theta_i|.
  \label{eq:bhalo}
\end{equation}

For each record we compute the surviving mass fraction of the cluster after its
encounters in the interval,
\begin{equation}
  s_i = \left[1 - f(\Delta v_{{\rm phys},i},\,b_i)\right]^{c_i},
  \label{eq:si}
\end{equation}
with $f$ given by the universal calibration of Equation~(\ref{eq:fcombined})
evaluated at the run's cluster parameters ($v_c$, $r_h$;
Section~\ref{sec:halo_sims}).  Records and halo-simulation particles are then
independently assigned to $100$ logarithmically spaced radial bins with edges
$r_j$ spanning $1$--$1000$ kpc in physical galactocentric radius.
Within each radial bin $j$ and snapshot interval $i$, the mean surviving fraction
per encountering cluster is
\begin{equation}
  \bar s_{i,j} = \left\langle s_i \right\rangle_{r\in[r_j,r_{j+1}]},
\end{equation}
and the fraction of clusters that encountered anything is
\begin{equation}
  \xi_{i,j} = \frac{E_{i,j}}{P_{i,j}},
\end{equation}
where $E_{i,j}$ and $P_{i,j}$ are the number of encounter records and the particle
count in bin $j$ during interval $i$.  The per-cluster surviving mass fraction in
bin $j$ after accumulating $N_{\rm snap}$ snapshots is then
\begin{equation}
  S_j = \prod_{i=1}^{N_{\rm snap}}\bar s_{i,j}^{\,\xi_{i,j}}.
  \label{eq:Sj}
\end{equation}
Note that Equation~(\ref{eq:Sj}) attributes mass loss to the radius at which each
encounter occurred; it is a per-radius mass-loss accounting rather than a Lagrangian
tracking of individual cluster trajectories.

\subsection{Radial Profile of Surviving Fraction}

\begin{figure*}[htbp]
  \centering
  \includegraphics[width=\linewidth]{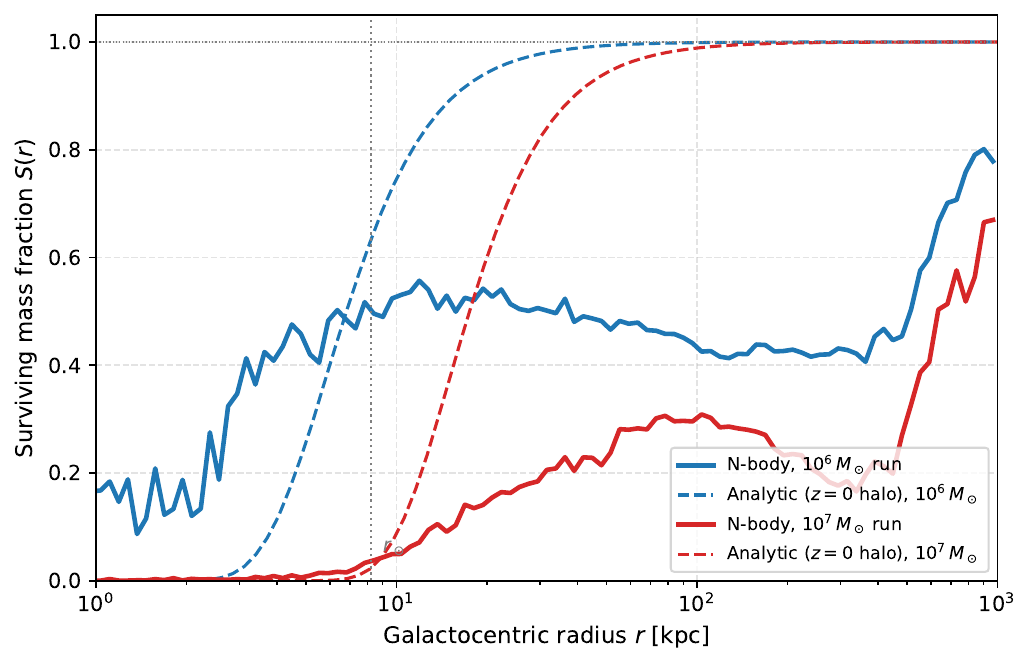}
  \caption{Cumulative surviving cluster mass fraction $S(r)$ as a function of
    physical galactocentric radius.
    \textit{Solid lines}: N-body encounter-integration result
    (Section~\ref{sec:survival}) for the $10^6\,M_\odot$ (blue) and
    $10^7\,M_\odot$ (red) runs.
    \textit{Dashed lines}: analytic estimate $S_{\rm an}=\exp(-\Gamma_{\rm eff}\,T)$
    with the measured $z=0$ NFW profile and $T=13.3\,\rm Gyr$.
    The analytic curves overestimate survival at large radii because they miss the
    dense, low-velocity encounters accumulated during halo assembly (see text).
    The vertical dotted line marks the Solar circle ($r_\odot=8.2\,\rm kpc$).}
  \label{fig:survival}
\end{figure*}

Figure~\ref{fig:survival} shows $S_j$ as a function of galactocentric radius for the
$10^6\,M_\odot$ (blue) and $10^7\,M_\odot$ (red) cluster simulations.  At the Solar
circle the surviving fraction is
\begin{equation}
  S(r_\odot) \simeq
  \begin{cases}
    0.50 & (10^6\,M_\odot), \\
    0.04 & (10^7\,M_\odot),
  \end{cases}
  \label{eq:ssun}
\end{equation}
declining further toward the Galactic center.  Although heavier $10^7\,M_\odot$
clusters have lower number density, their much larger cross-sections
($\sigma\propto r_h^2\propto M_{\rm cl}^{5/3}$) dominate, and they are almost
entirely disrupted throughout the halo.

Figure~\ref{fig:survival} also shows the analytic estimate
$S_{\rm an}(r)=\exp\!\bigl(-\Gamma_{\rm eff}(r)\,T\bigr)$, where $\Gamma_{\rm eff}$
is the $f$-weighted disruption rate --- the integrand of
Section~\ref{sec:analytic} multiplied by $f(v_{\rm rel},b)$ and integrated over
impact parameters $b\le r_{\log}$ --- evaluated for the present-day ($z=0$) NFW fit
and the same cluster parameters as the simulation, with $T=13.3\,\rm Gyr$.  The two
estimates agree reasonably near the Solar circle ($S_{\rm an}(r_\odot)\simeq0.63$
vs.\ $0.50$ for $10^6\,M_\odot$; $0.02$ vs.\ $0.04$ for $10^7\,M_\odot$), but
diverge strongly at larger radii: the analytic survival rises to unity beyond
${\sim}30$--$50\,\rm kpc$, whereas the N-body result plateaus at $S\simeq0.45$
($10^6\,M_\odot$) and $S\simeq0.2$--$0.3$ ($10^7\,M_\odot$) out to several hundred
kpc, recovering toward unity only in the distant, never-collapsed outskirts.

\begin{figure*}[htbp]
  \centering
  \includegraphics[width=\linewidth]{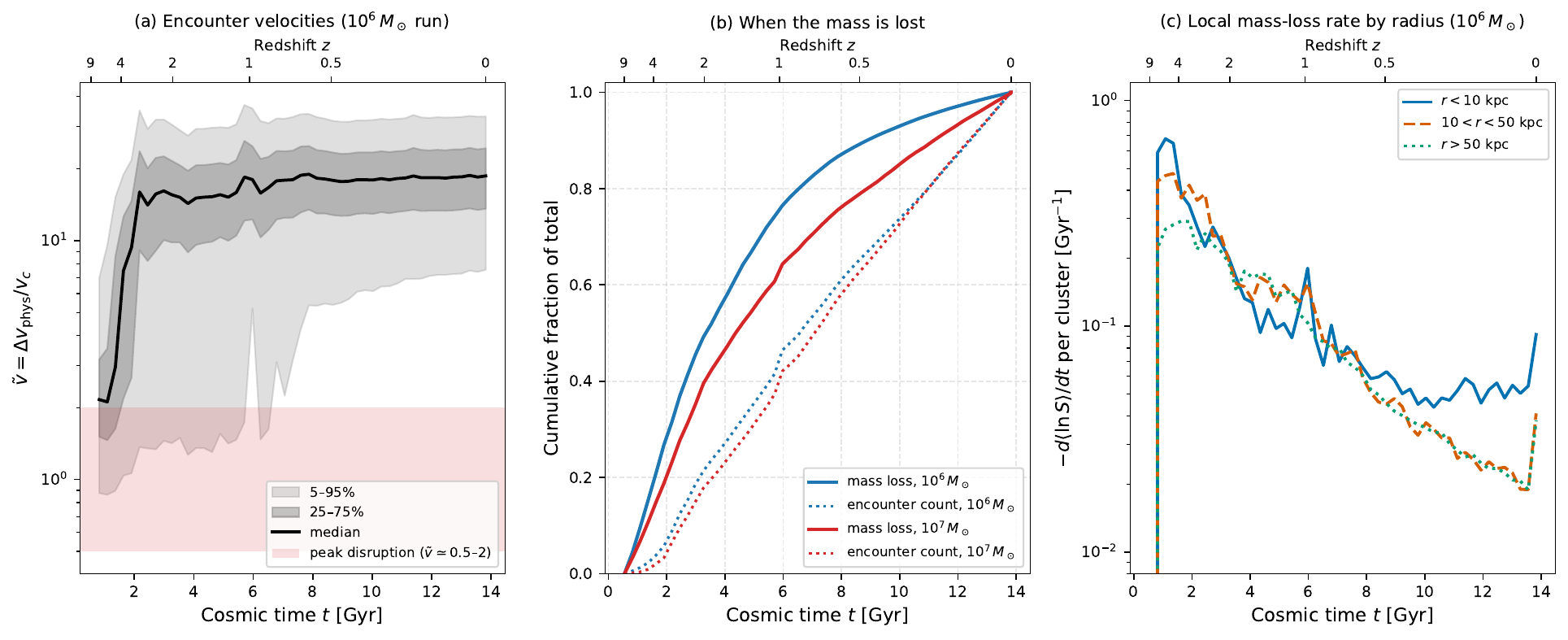}
  \caption{When the cluster mass is lost.
    \textit{Panel (a)}: count-weighted percentiles of the physical encounter
    velocity in the $10^6\,M_\odot$ run, in units of the cluster internal velocity
    scale $v_c$, as a function of cosmic time.  The shaded horizontal band marks
    the peak-disruption range $\tilde v\simeq0.5$--$2$ (Figure~\ref{fig:fv}).
    \textit{Panel (b)}: cumulative fraction of the total integrated mass loss
    (solid) and of the total encounter count (dotted) for both runs.  Mass loss
    accumulates much earlier than the encounters themselves: half of the damage is done by
    $z\approx2$ ($10^6\,M_\odot$; $z\approx1.5$ for $10^7\,M_\odot$), while half
    of the encounters occur only after $z\approx0.85$.
    \textit{Panel (c)}: per-cluster mass-loss rate $-d\langle\ln S\rangle/dt$ in
    three bins of the radius at which the encounters occur ($10^6\,M_\odot$ run):
    the rate declines by an order of magnitude from $z\approx4$ to $z=0$ at every
    radius.}
  \label{fig:history}
\end{figure*}

The discrepancy has a straightforward physical explanation, which
Figure~\ref{fig:history} demonstrates quantitatively.  During the halo's
hierarchical assembly the proto-halo consists of cold, dense substructures in which
clusters meet at relative velocities comparable to their internal velocity scale:
at $z\gtrsim4$ the count-weighted median encounter velocity is
$\tilde v\simeq2$--$3$, with a quarter of all encounters inside the peak-disruption
band $\tilde v\simeq0.5$--$2$ (panel a).  As the halo virializes, the median climbs
to $\tilde v\simeq17$ ($\Delta v\simeq230\,\rm km\,s^{-1}$), deep in the harmless
fast-fly-by regime.  Encounters themselves keep occurring at a nearly constant pace
--- half of the total encounter \emph{count} accumulates only by $z\approx0.85$ ---
but they stop removing mass: half of the total integrated mass \emph{loss} is
already in place by $z\approx2.0$ ($t\approx3.3\,\rm Gyr$) for $10^6\,M_\odot$
clusters and by $z\approx1.5$ for $10^7\,M_\odot$ clusters (panel b).
Correspondingly, the per-cluster mass-loss rate declines by an order of magnitude
between $z\approx4$ and $z=0$ at every radius (panel c).  The analytic formula,
which applies the present-day rate of a smooth, virialized halo over the full
Hubble time, misses this early channel entirely.  Because
Equation~(\ref{eq:Sj}) attributes mass loss to the radius at which it occurred, the
assembly-epoch damage appears as the extended plateau at $r\gtrsim10\,\rm kpc$; the
cluster population found at these radii today passed through dense progenitors
earlier in its history.  Beyond $R_{200}$ this cold-encounter channel persists to
the present day inside infalling substructures
(Section~\ref{sec:discussion}).  The same assembly signature is visible in the
encounter-rate comparison (Figure~\ref{fig:gamma_r}) as the N-body excess over the
smooth-halo prediction at large radii.  The analytic $S_{\rm an}(r)$ should
therefore be read as an upper bound on survival outside the inner halo, and the
N-body $S(r)$ as the more reliable result.

\section{Line-of-Sight Smooth Fraction}
\label{sec:los}

We define the \textit{smooth fraction} as
$f_{\rm sm}(r)\equiv 1-S(r)$, the fraction of PBH mass that has been stripped from
clusters and is now distributed as a diffuse, unclustered component.  The
observationally relevant quantity for microlensing is the DM-mass-weighted average of
$f_{\rm sm}$ along the line of sight from the Sun to the source,
\begin{equation}
  \langle f_{\rm sm}\rangle_{\rm LOS} =
  \frac{\displaystyle\int_0^{d_{\rm src}}\rho_{\rm DM}(R(s))\,f_{\rm sm}(R(s))\,ds}
  {\displaystyle\int_0^{d_{\rm src}}\rho_{\rm DM}(R(s))\,ds},
  \label{eq:los}
\end{equation}
where $s$ is the heliocentric distance along the line of sight, and
$R(s) = [R_\odot^2 + s^2\cos^2 b - 2R_\odot s\cos b\cos\ell + s^2\sin^2 b]^{1/2}$
is the Galactocentric distance at position $s$, with $R_\odot=8.2\,\rm kpc$; here
$(\ell,\,b)$ are the Galactic longitude and latitude of the source (this $b$, used
only in the present section, is unrelated to the impact parameter).

We evaluate Equation~(\ref{eq:los}) toward the Large Magellanic Cloud
($\ell=280\fdg5$, $b=-32\fdg9$, $d_{\rm src}=49.5\,\rm kpc$) and the Small Magellanic Cloud
($\ell=302\fdg8$, $b=-44\fdg3$, $d_{\rm src}=62.0\,\rm kpc$),
using $2000$ uniformly spaced steps in $s$.

In Equation~(\ref{eq:los}), $\rho_{\rm DM}(R(s))$ is evaluated from the analytic NFW
profile with the measured parameters of Equation~(\ref{eq:nfw_params}); this provides
a smooth, consistent weight along the sightline.

\begin{figure*}[htbp]
  \centering
  \includegraphics[width=\linewidth]{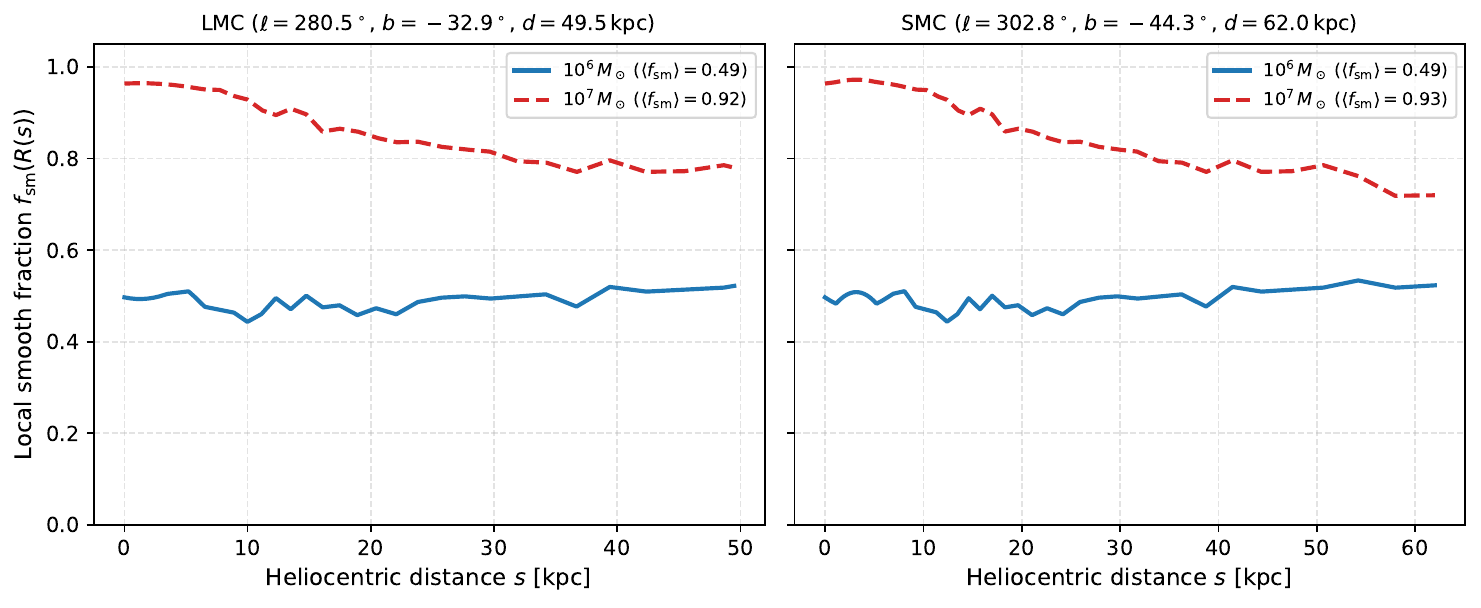}
  \caption{Local smooth fraction $f_{\rm sm}(R(s))=1-S(R(s))$ as a function of
    heliocentric distance $s$ along the LMC (left) and SMC (right) sightlines,
    from the N-body survival profiles of Figure~\ref{fig:survival}:
    $10^6\,M_\odot$ clusters (blue solid) and $10^7\,M_\odot$ clusters (red
    dashed).  Legends give the DM-mass-weighted averages
    $\langle f_{\rm sm}\rangle_{\rm LOS}$ (Table~\ref{tab:los}).}
  \label{fig:los}
\end{figure*}

Figure~\ref{fig:los} shows the local smooth fraction $f_{\rm sm}(R(s))$ as a function
of heliocentric distance for both sightlines and both cluster mass scales.  Because
the N-body survival profile is nearly flat over the radial range probed by the
sightlines ($R\sim8$--$60\,\rm kpc$), the local smooth fraction is nearly uniform
along the column: $f_{\rm sm}\simeq0.45$--$0.52$ for $10^6\,M_\odot$ clusters and
$f_{\rm sm}\simeq0.72$--$0.97$ for $10^7\,M_\odot$ clusters, the latter declining
slowly outward.

Table~\ref{tab:los} presents the DM-mass-weighted $\langle f_{\rm sm}\rangle_{\rm LOS}$
for both cluster mass scales.  The LMC sightline passes predominantly
through the inner halo (Galactocentric distances $R\sim 8$--$50\,\rm kpc$), where cluster
disruption is most active, while the SMC probes a comparable but somewhat more extended
column; the resulting averages are nearly identical for the two sightlines.

\begin{table}[ht]
  \caption{\label{tab:los}DM-mass-weighted smooth fraction $\langle f_{\rm sm}\rangle_{\rm LOS}$
    along the LMC and SMC sightlines, from the N-body survival profiles of the
    MW-like halo simulations.}
  \begin{ruledtabular}
  \begin{tabular}{lcc}
    Sightline & $M_{\rm cl}=10^6\,M_\odot$ & $M_{\rm cl}=10^7\,M_\odot$ \\
    \hline
    LMC & $0.486$ & $0.917$ \\
    SMC & $0.489$ & $0.926$ \\
  \end{tabular}
  \end{ruledtabular}
\end{table}

A larger smooth fraction implies that more PBH mass resides in individual,
freely moving black holes rather than in bound clusters, and that the microlensing optical
depth and event-duration distribution are correspondingly closer to the predictions for
an unclustered population of $30\,M_\odot$ compact objects.  Conversely, a small smooth
fraction indicates that clustering must be explicitly modelled when applying microlensing
constraints.  We find $\langle f_{\rm sm}\rangle\simeq0.49$ for $10^6\,M_\odot$
clusters --- an almost even split between the clustered and smooth components ---
and $\simeq0.92$ for $10^7\,M_\odot$ clusters, which are nearly fully disrupted
along these sightlines.

\section{Discussion}
\label{sec:discussion}

At the Solar circle, the late-time ($z\lesssim0.2$) N-body encounter rate agrees with
the analytic model evaluated at the simulation's own parameters to within $\sim10\%$
for the $10^6\,M_\odot$ run and $\sim40\%$ for the $10^7\,M_\odot$ run
(Figure~\ref{fig:gamma_r}).  Two systematic departures frame this agreement.  In the
innermost few kpc the N-body rate falls below the analytic curve, most likely
because the smooth-density, Maxwellian-velocity assumptions of the analytic model
degrade toward the halo center.  Whatever its origin, the deficit occurs where the
encounter velocities are highest and is therefore harmless for the survival
analysis: encounters at those velocities carry $\tilde f\approx0$.
Beyond ${\sim}30\,\rm kpc$ the N-body
rate increasingly exceeds the smooth-halo prediction (up to an order of magnitude at
$100\,\rm kpc$): the analytic model assumes an unclustered NFW medium, whereas the
real outer halo is rich in substructure, within which the local density --- and hence
the encounter rate --- far exceeds the spherical average.

Substructure changes not only the frequency of encounters but also their
velocities, and the two effects dominate in different regions.  Within $R_{200}$
the late-time median encounter velocity stays close to the smooth-halo Maxwellian
expectation ($250$--$275\,\rm km\,s^{-1}$ at $10$--$200\,\rm kpc$), so the rate
excess there is mainly a density effect.  Beyond $R_{200}$, however, encounters
occur between clusters bound inside infalling groups and filament clumps, whose
internal dispersions are far below the halo virial velocity: the late-time median
relative velocity at $200$--$400\,\rm kpc$ collapses to
${\simeq}18\,\rm km\,s^{-1}$ --- an order of magnitude below the smooth-halo
expectation, at the disruption peak, $\tilde v\approx1.3$.  This cold
substructure channel is the same one that dominates the accumulated disruption
during halo assembly (Section~\ref{sec:survival}, Figure~\ref{fig:history});
at $z=0$ it survives only outside the virialized region, which is why $S(r)$
remains depressed out to several hundred kpc and recovers only in the
never-collapsed outskirts.

Because the simulated halo is consistent with the Milky Way, the rates of
Equation~(\ref{eq:gammanumerical}) apply directly to the Galaxy.  We note that
the halo mass, $M_{200}\simeq7.8\times10^{11}\,M_\odot$, lies at the lower end of
current MW estimates; for a halo at the upper end
($M_{200}\simeq1.3\times10^{12}\,M_\odot$ at fixed concentration), the local density
and velocity scale at the Solar circle are higher and the rates in
Equation~(\ref{eq:gammanumerical}) rise by ${\sim}50\%$, without changing
any qualitative conclusion.

The cluster size $r_h$ introduces some additional uncertainty.  We adopt the
formation-model values from Equation~(\ref{eq:rh}): $r_h\simeq25\,\rm pc$ and
$168\,\rm pc$ for the $10^6$ and $10^7\,M_\odot$ cases.  A key feature of the Carr
scaling $r_h\propto M_{\rm cl}^{5/6}$ is that the collision cross-section grows as
$\sigma\propto r_h^2\propto M_{\rm cl}^{5/3}$, while the cluster number density falls
as $n_{\rm cl}\propto M_{\rm cl}^{-1}$; the net collision rate therefore scales as
$\Gamma\propto M_{\rm cl}^{2/3}$, so \emph{more massive clusters collide more
frequently}.  Varying $r_h$ by a factor of two shifts the halo-averaged rate by a
factor of four but does not change the qualitative conclusion that $10^7\,M_\odot$
clusters are fully disrupted.

Several effects make our disruption estimates conservative.  The halo simulations
log encounters only within $r_{\log}\simeq4.3\,r_h$; because the impact-parameter
suppression declines slowly, $g\propto(\tilde b/7.4)^{-2.2}$ at large $\tilde b$,
more distant tidal encounters contribute additional cumulative mass loss that our
accounting omits.  Our simulations also include only the dark-matter component, so
dynamical friction from
baryons (stellar disk, bulge, and gas) is absent.  Baryonic friction would cause
massive clusters to spiral inward on timescales of several Gyr; clusters that migrate
to smaller radii encounter higher densities and lose more mass.  On both counts the
present $S(r)$ is an upper bound on survival --- and $f_{\rm sm}$ a lower bound on
the smooth fraction.  At the low-velocity end
($\tilde v \lesssim \tilde v_{\rm b}$), binary collisions may end in mergers rather
than disruption.  Our slowest runs ($\tilde v\approx0.02$) show a residual escape
fraction of $3$--$4\%$,
likely from three-body ejections during violent relaxation; the merger product is a more
massive, presumably more compact cluster, whose subsequent evolution we do not track.

Our accounting also treats every cluster's parameters as fixed: each encounter is
evaluated with the initial $(M_{\rm cl},\,r_h)$, regardless of the mass already
lost.  In reality, a cluster that has lost a large fraction of its mass presents a
smaller cross-section and collides less often, which slows further disruption; at
the same time, if stripping leaves a more compact remnant, its higher $v_c$ moves
halo encounters closer to the disruption peak and makes each of them more damaging.
The stripped PBHs are not passive either: they build up the smooth component
through which the surviving clusters move, so dynamical friction against this
background slowly drags clusters toward the Galactic center --- into regions of
higher encounter rates --- and gradually modifies the halo profile itself.  These
couplings lie beyond the independent-encounter approximation used here and call
for simulations that follow the cluster population self-consistently.

Turning to the observational implications, the smooth fractions in Table~\ref{tab:los}
are the central result for microlensing.  Along the LMC sightline, even the most
optimistic case --- $10^6\,M_\odot$ clusters --- gives
$\langle f_{\rm sm}\rangle\simeq0.49$: half the PBH mass is already in
free-flying individual black holes, subject to standard microlensing constraints as if
the population were unclustered.  For $10^7\,M_\odot$ clusters the fraction rises to
$0.92$.  The original appeal of clustering as a way to evade microlensing limits is
therefore only partially realized: a substantial fraction of the inner-halo cluster
population is disrupted over the Galaxy's lifetime regardless of initial cluster mass.
What clustering does achieve, for the remaining bound fraction $1-\langle f_{\rm sm}\rangle$,
is to shift event timescales and change the lensing cross-section relative to individual
PBHs.  The practical consequence is that standard exclusion limits from surveys such as
EROS or OGLE apply directly to the smooth component ($f_{\rm sm}\times f_{\rm PBH}$ of
the total DM), while the clustered remainder requires a separate lensing model.
PBH clustering does not remove the microlensing constraints; it partitions them between
two components with different observational signatures.

We stress that $f_{\rm sm}$ as defined here is a strict lower bound on the
\emph{effectively unclustered} PBH population for microlensing purposes.  Our model
treats each surviving cluster as a monolithic, opaque object: all mass within it is
assigned to the ``clustered'' component and assumed to produce non-standard lensing
signatures.  \citet{Toshchenko2025} showed, via ray-tracing simulations of PBH cluster
lensing, that this assumption is overly conservative: even within intact clusters the
low-density periphery produces microlensing light curves indistinguishable from isolated
point-lens events, with the point-lens point-source event fraction approaching unity at the cluster edge.
Accounting for this peripheral contribution would raise the effectively smooth fraction
above the values in Table~\ref{tab:los}, further reinforcing the conclusion that
standard microlensing limits constrain a substantial portion of any putative PBH dark
matter population.

A further caveat concerns the internal composition of the clusters when PBHs make up
only a fraction $f_{\rm PBH}<1$ of the dark matter.  In that case each PBH accretes
a dense ``dress'' of ordinary CDM, with a $\rho\propto r^{-9/4}$ density profile and
a mass growing roughly in proportion to the scale factor
\citep{Tkachev2020, Pilipenko2022}.  Clusters would then consist of dressed PBHs
embedded in an additional smooth CDM component.  The extra mass deepens the cluster
potential and raises the internal velocity scale $v_c$, making clusters effectively
more compact and more resistant to disruption; at the same time, heavier clusters
collide more frequently ($\Gamma\propto M_{\rm cl}^{2/3}$ under the Carr scaling),
and the dresses themselves add a dissipative channel to each encounter.  The net
effect on the survival fraction is not obvious and warrants a dedicated study.

\section{Conclusions}
\label{sec:conclusions}

We have presented a combined analytic and N-body study of cluster-cluster collisional
disruption of PBH clusters in a Milky Way-like dark matter halo, simulated from
$z=9$ to $z=0$ with every simulation particle representing one PBH cluster.

The re-simulated halo has $M_{200}\simeq7.8\times10^{11}\,M_\odot$ and
$c_{200}\simeq11$, measured directly from the $z=0$ snapshot, consistent with
Milky Way values.  With these parameters the analytic local per-cluster encounter
rate at the Solar circle is
$\Gamma(r_\odot)\simeq 2.9\times10^{-3}\,\rm Myr^{-1}$
($\tau_{\rm coll}\simeq 350\,\rm Myr$) for $10^6\,M_\odot$ clusters and
$1.2\times10^{-2}\,\rm Myr^{-1}$ ($\tau_{\rm coll}\simeq 82\,\rm Myr$) for
$10^7\,M_\odot$ clusters --- both far shorter than the Hubble time --- with the rate
rising steeply toward the Galactic center as $\Gamma\propto\rho_{\rm NFW}(r)$.  The
late-time N-body encounter rate confirms the analytic model at the Solar circle to
within $\sim10\%$--$40\%$; in the outer halo the simulation exceeds the smooth-halo
prediction because of substructure.

Binary N-body collisions yield a peak escape fraction of $\sim 22\%$ for head-on
encounters at $v_{\rm rel}\simeq(0.9$--$1.2)\,v_c$, where
$v_c=\sqrt{GM_{\rm cl}/r_h}$ is the cluster internal velocity scale.  The calibration
is a universal dimensionless function
$\tilde f(\tilde v,\tilde b)=\tilde f_v(\tilde v)\,g(\tilde b)$ of
$\tilde v=v_{\rm rel}/v_c$ and $\tilde b=b/r_h$, exact for any cluster mass and
radius at fixed profile shape and verified by direct simulations at the Carr radius.
For the Carr formation radii the disruption peak corresponds to
$v_{\rm rel}\approx 12$--$16\,\rm km\,s^{-1}$ ($10^6\,M_\odot$) and
$15$--$20\,\rm km\,s^{-1}$ ($10^7\,M_\odot$), well below typical halo encounter
velocities: individual halo encounters remove $\lesssim10^{-2}$ of a cluster's mass,
and disruption accumulates over many weak encounters.

Integrating the recorded encounter histories with the universal calibration, the
surviving cluster mass fraction at the Solar circle is $S(r_\odot)\simeq0.50$ for
$10^6\,M_\odot$ clusters and $\simeq0.04$ for $10^7\,M_\odot$ clusters, declining
further toward the Galactic center.  A $z=0$ analytic estimate
$S_{\rm an}=\exp(-\Gamma_{\rm eff}\,T)$ reproduces the N-body result near the Solar
circle but overestimates survival at larger radii, because most of the accumulated
damage is inflicted during hierarchical assembly, when clusters met inside cold,
dense progenitors at relative velocities near the disruption peak
($\tilde v\sim1$): half of the total mass loss is in place by $z\approx2$, and the
per-cluster mass-loss rate declines by an order of magnitude between $z\approx4$
and $z=0$ at every radius --- a channel entirely absent from any present-day
smooth-halo estimate.

The DM-mass-weighted smooth fraction along the LMC and SMC sightlines is
$\langle f_{\rm sm}\rangle\simeq0.49$ for $10^6\,M_\odot$ clusters and
$\simeq0.92$ for $10^7\,M_\odot$ clusters, and directly quantifies what fraction of
the PBH dark matter along microlensing sightlines should be modelled as individual
compact objects versus extended cluster lenses.

Our results establish that collisional disruption is a substantial process over the
lifetime of the Galaxy for cluster masses in the range $10^6$--$10^7\,M_\odot$: at
least half of the initially clustered PBH mass along the Magellanic sightlines is
released into a smooth component.  A proper reanalysis of existing
microlensing data should account for both the clustered and smooth components of PBH
dark matter, weighted by the radially varying $f_{\rm sm}(r)$ derived here.

Future work will extend the binary collision grid to include clusters of different
masses and density profiles (addressing the head-on vs. unequal-mass case), incorporate
dynamical friction in the halo evolution and the ordinary-CDM dresses of individual
PBHs relevant for $f_{\rm PBH}<1$ \citep{Tkachev2020, Pilipenko2022}, and apply the
resulting smooth-fraction
profiles to derive revised, clustered-population constraints from EROS, OGLE, and
Subaru HSC data.

\begin{acknowledgments}
We are eternally indebted to P. B. Ivanov for many illuminating discussions and comments.
\end{acknowledgments}

\bibliography{refs}


\appendix

\section{Two-Body Integration for Simulation Duration}
\label{app:tmax}

The simulation duration for each binary collision run is set by integrating the reduced
two-body problem in the orbital plane.  The relative coordinate
$\mathbf{r} = \mathbf{r}_2 - \mathbf{r}_1$ evolves under
\begin{equation}
  \ddot{\mathbf{r}} = -\frac{G\cdot 2M_{\rm cl}}{(|\mathbf{r}|^2 + \epsilon^2)^{3/2}}\,\mathbf{r},
\end{equation}
starting from the initial separation $r_0=100\,\rm pc$, with a Plummer softening
$\epsilon=50\,\rm pc$ of order the cluster-pair scale.
Integration proceeds with a timestep $\Delta t=10^{-3}$ code units until the separation
exceeds $r_{\rm esc}=200\,\rm pc$ (and the clusters are separating, $\dot r>0$), at which
point $t_{\rm esc}$ is returned as \texttt{TimeMax}.  For gravitationally bound pairs
($E_{\rm tot}<0$) a maximum integration time of $8\,\rm Myr$ is imposed.

\end{document}